\documentclass[twocolumn,superscriptaddress, floatfix,amsmath,amssymb,nofootinbib,prd]{revtex4-1}
\usepackage[english]{babel}
\usepackage{graphicx}
\usepackage{dcolumn}
\usepackage{bm}
\usepackage{verbatim}
\usepackage{mathrsfs}
\usepackage{mathtools}
\usepackage[normalem]{ulem}
\usepackage{natbib}
\usepackage{color}
\usepackage{hyperref}
\usepackage[dvipsnames]{xcolor}
\usepackage[utf8]{inputenc}

\pdfoutput=1

\renewcommand{\Re}{\operatorname{Re}}
\renewcommand{\Im}{\operatorname{Im}}
\newcommand{\ii}{\mathrm{i}}
\newcommand{\dd}{\mathrm{d}}

\begin{document}

\title{New techniques for entanglement harvesting in flat and curved spacetimes}
\author{Keith K. Ng}
\affiliation{Department of Physics and Astronomy, University of Waterloo, Waterloo, ON, N2L 3G1, Canada}
\author{Robert B. Mann}
\affiliation{Department. Physics and Astronomy, University of Waterloo, Waterloo, ON, N2L 3G1, Canada}
\affiliation{Institute for Quantum Computing, University of Waterloo, Waterloo, Ontario, N2L 3G1, Canada}
\affiliation{Perimeter Institute for Theoretical Physics, Waterloo, ON, N2L 2Y5, Canada}
\author{Eduardo Mart\'{i}n-Mart\'{i}nez}
\affiliation{Department of Applied Mathematics, University of Waterloo, Waterloo, ON, N2L 3G1, Canada}
\affiliation{Institute for Quantum Computing, University of Waterloo, Waterloo, Ontario, N2L 3G1, Canada}
\affiliation{Perimeter Institute for Theoretical Physics, Waterloo, ON, N2L 2Y5, Canada}

\begin{abstract}
 We present a new technique for computing entanglement harvesting with Unruh-DeWitt  particle detectors. The method is particularly useful in cases where analytic solutions are rare and the Wightman function is known only via its mode expansion for which numerical integration can become very expensive. By exploiting the conjugate symmetry of the Wightman function, we may split the integral into parts dependent on the commutator and anti-commutator of the field. In cases where the commutator vanishes, such as spacelike separation, or timelike separation if the strong Huygens principle holds, we then show that the entangling term of the bipartite density matrix can be expressed in terms of the much simpler \textit{mutual information term}. For the vacuum state, this can be translated into a simple Fourier transform, and thus a single sum over modes, simplifying the procurement of closed expressions. We demonstrate this for Minkowski space, finding an analytical solution where none was previously known.
\end{abstract}

\maketitle

\section{Introduction}

The quantum vacuum has proved to be a very interesting setting for studying foundational ideas in physics. A free quantum field in the vacuum state has for some time been known to contain correlations between time- and spacelike-separated regions \cite{Algebra1,Algebra2}.  More recently it has been shown that such correlations can, in principle, be extracted from the vacuum, a phenomenon referred to as entanglement harvesting \cite{VALENTINI1991321,Reznik:2002fz}.

The theoretical tools for extracting vacuum entanglement are referred to as detector models, the best-known of which is called the Unruh-DeWitt model.  Under certain conditions,  Unruh-DeWitt detectors  provide a good approximation to the light-matter interaction  (see section II of \cite{Pozas2016} and, e.g.,  \cite{PhysRevD.87.064038,PhysRevA.89.033835}), raising the possibility that atomic qubits could harvest entanglement from the electromagnetic vacuum for use later as a resource for quantum information tasks \cite{Martinez2013a}.   

However, calculating the amount of entanglement harvested is quite challenging. In order to to describe the quantum state of the field, one uses the Wightman function, a two-point correlator. In the simplest cases, such as in Minkowski space, the Wightman function may be described as a function of the geodesic distance between two points, allowing for analytic results \cite{Pozas-Kerstjens:2015gta,Sachs2017,Koga:2018the} However, in more general cases, this expression is not available; the Wightman often must be expressed in terms of a sum over field modes.

This latter case is  generally  more complicated to deal with. Even within perturbation theory,  to evaluate entanglement between two detectors interacting with the vacuum, one must evaluate several integrals over the proper time of the detectors of the Wightman function pulled-back to the detectors' trajectory. Since the Wightman function is often not known analytically, we must do the integration numerically. In the case where the Wightman is only available as a mode expansion, then, we must integrate multiple times: twice over time, and once more over momenta modes. This leads to a much higher computational complexity.

Naively, one might expect this case to be amenable to Fourier transformation, at least in the simplest physical configurations. If the spacetime has a timelike Killing vector, and the detector remains stationary, the transition rate of the detector may indeed be expressed this way, as we found in \cite{Ng:2016hzn}.  In the general case this is a complicated endeavour, and computation of terms in the density matrix associated with non-local correlations cannot be carried out in this way.

In this paper we  show that in certain special cases, the Fourier form may be recovered.  It then becomes possible to do the time integrations analytically, and then we only need sum over mode energies. This is a significant reduction in computational complexity, and offers some conceptual clarity. We believe this result may be applied quite generally to calculations of entanglement harvesting, and to great effect.
 
\section{Basic formalism}

Consider two Unruh-DeWitt detectors A,B: that is, two two-level systems of proper gap $\Omega_I$ with coupling $\lambda_I$ to a scalar field. We will \textit{not} require them to be physically identical. These detectors are switched on and off according to $\chi_I(\tau_I)$, with respect to their own proper time. With respect to some common coordinate time, the Hamiltonian density of the interaction with the field may be described by 
\begin{align}
    \hat H_I\!&=\!\sum_{I=A,B}\!\!\!\!\lambda_I \left|\frac{\partial(\tau_I,\bm{\xi}_I)}{\partial(t,\bm{x})}\right|\chi_I(\tau_I(t,\bm x))
    \hat{\mu}_I(\tau_I(t,\bm{x}))\nonumber\\
    &\times F_I(\bm{\xi}_I(t,\bm{x}))\hat{\Phi}(t,\bm{x}),
\end{align}
where $\hat\mu_I(\tau_I)=e^{\ii\Omega_I\tau_I}\hat\sigma_I^++e^{-\ii\Omega_I\tau_I}\hat\sigma_I^-$ is the monopole operator, $F_I(\bm{\xi}_I(t,\bm{x}))$ describes the spatial smearing of detector $I$ at a given proper time, and $\left|\frac{\partial(\tau_I,\bm{\xi}_I)}{\partial(t,\bm{x})}\right|$ describes the Jacobian of the transformation to the detector's smearing frame  $(\tau_I,\bm{\xi}_I)$ \cite{Martin-Martinez:2018gzb}. (As other authors have noted \cite{PhysRevA.81.012330,Martin-Martinez:2015psa}, a non-pointlike detector may cause certain issues with the causality of the model  if not handled carefully. We will not discuss this much further at this time, but merely note that the pointlike case may also be expressed in this form and refer to the discussion in \cite{Martin-Martinez:2015psa} and \cite{Martin-Martinez:2018gzb}.)

Using this interaction Hamiltonian, we may calculate the Dyson expansion of the unitary evolution operator. Expanding in powers of $\lambda_A, \lambda_B$,  we find
\begin{align}
    \hat U(t,t_0)=&\sum_{n=0}^\infty \hat U_n,\\
    \hat U_n(t,t_0)=&\frac{(-\ii)^n}{n!}\int_{t_0}^t dt_1 \int_{t_0}^t dt_2 \cdots \int_{t_0}^t dt_n \nonumber\\
    &\mathcal{T}\left(\hat H_I(t_1)\hat H_I(t_2)\cdots \hat H_I(t_n)\right)
\end{align}
with $U_0$ is simply the identity operator. Finally, since we only have access to the detector states, we trace over  the field degrees of freedom. Thus, the final state of the detectors may be described as
\begin{equation}
    \hat\rho_{AB}(t)=\sum_{n=0}^\infty \sum_{n'=0}^\infty \mathrm{Tr}_\phi [\hat U_n(t,t_0)\hat\rho_0 \hat U_{n'}^\dagger(t,t_0)],
\end{equation}
where $\hat\rho_0$ describes the joint initial state of the detectors and field. For our purposes, we will initialize the detectors in their ground state.
Note the double sum: for instance, the second-order contributions to the state are described by
\begin{align}
     \hat\rho_{AB}^{(2)}(t)=\mathrm{Tr}_\phi [&\hat U_2(t,t_0)\hat\rho(t_0)\hat U_0^\dagger(t,t_0)\nonumber\\
     &+\hat U_1(t,t_0)\hat\rho(t_0)\hat U_1^\dagger(t,t_0)\nonumber\\
     &+\hat U_0(t,t_0)\hat\rho(t_0)\hat U_2^\dagger(t,t_0)].
\end{align}
This is necessary to maintain normalization to all orders. The $\hat U_2\hat\rho_0 \hat U_0^\dagger+\hat U_0\hat\rho_0 \hat U_2^\dagger$ term, for instance, includes a $\lambda_A^2$ term, which reduces the probability detector A is observed in its ground state by the same amount that $\hat U_1\hat\rho_0 \hat U_1^\dagger$ increases the probability it is observed in its excited state. Of course, the `mixed' terms proportional to $\lambda_A\lambda_B$ also contribute to different types of correlations in the final density matrix \cite{Martin-Martinez:2015psa}.

The effect of the trace over the field is to make each term proportional to an $n$-point correlator of the field. For field states in which odd-$n$ correlators vanish, such as the vacuum state, Fock states, and free thermal states, the leading contribution to the state is therefore $\hat\rho_{AB}^{(2)}.$ Let us define a basis for the two-detector state as
\begin{equation}
    \begin{matrix}
        |g_A g_B\rangle=(1,0,0,0)^\dagger & |e_A g_B\rangle=(0,1,0,0)^\dagger\\
        |g_A e_B\rangle=(0,0,1,0)^\dagger & |e_A e_B\rangle=(0,0,0,1)^\dagger.
    \end{matrix}
\end{equation}
It can then be shown (e.g. \cite{Sachs2017}) that the terms of the density matrix can be written as an integral transform of the Wightman function, \mbox{$W(t,\bm{x},t',\bm{x'})=\langle \Psi | \hat\Phi(t,\bm{x}) \hat\Phi(t',\bm{x'}) | \Psi \rangle$}. In this way, the Wightman function characterizes the observables of this configuration to leading order.
Specifically, quantizing with respect to the Killing time $t$, we find \cite{Pozas-Kerstjens:2015gta}
\begin{equation}\label{rho2}
\hat{\rho}_{AB,t}=
\begin{bmatrix}
    1-\mathcal{L}_{AA}-\mathcal{L}_{BB} & 0 & 0 & \mathcal{M}^* \\
    0 & \mathcal{L}_{AA} & \mathcal{L}_{AB} & 0 \\
    0 & \mathcal{L}_{BA} & \mathcal{L}_{BB} & 0 \\
    \mathcal{M} & 0 & 0 & 0
\end{bmatrix}
+O(\lambda_I^4),
\end{equation}
where
\begin{align}
\mathcal{M}=&-\lambda_A \lambda_B \int_{-\infty}^{\infty} \dd t \int_{-\infty}^{t}\dd t' \int \dd^n\!\bm{x}\int \dd^n\!\bm{x'}\nonumber\\ &\times \sqrt{g(t,\bm x)g(t',\bm x')} \mathcal{M}(t,\bm{x},t',\bm{x'})W(t,\bm{x},t',\bm{x'})
\label{Mintegral}\\
\mathcal{L}_{IJ}=&\lambda_I \lambda_J \int_{-\infty}^{\infty} \dd t \int_{-\infty}^{\infty}\dd t' \int \dd^n\!\bm{x}\int \dd^n\!\bm{x'}\nonumber\\
&\times  \sqrt{g(t,\bm x)g(t',\bm x')} \mathcal{L}_I(t,\bm{x})\mathcal{L}_J^*(t',\bm{x'})W(t,\bm{x},t',\bm{x'})\label{LLintegral}
\end{align}
and 
\begin{align}
&\mathcal{L}_I(t,\bm{x})=\chi_I(\tau_I(t,\bm x))\left|\frac{\partial(\tau_I,\bm{\xi}_I)}{\partial(t,\bm{x})}\right|F_I(t,\bm{x})e^{\ii\Omega_I \tau_I(t,\bm x)} 
 \label{Lintegrand}
\\
&\mathcal{M}(t,\bm{x},t',\bm{x'})=\mathcal{L}_A(t,\bm{x})\mathcal{L}_B(t',\bm{x'})+\mathcal{L}_A(t',\bm{x'})\mathcal{L}_B(t,\bm{x})\label{Mintegrand}
\end{align}
where $I = A,B$.

Each term in \eqref{rho2} has a physical interpretation: the quantity $L_{JJ}$ corresponds to the excitation probability of detector J, while $L_{JI}$ conveys the shared information between the detectors. The term $\mathcal{M}$ in 
equation \eqref{Mintegral} is the key object of interest. This term is  associated with the entanglement harvested by the detectors, and has been traditionally called the \textit{entangling term}; however, the integration limits impose time ordering, which prevents its interpretation as a straightforward Fourier transform. It also leads to some difficulties with performing the integration; see \cite{Pozas-Kerstjens:2015gta} for a worked example. 

We turn now to investigating  how to reexpress this integral in more easily understood forms.

\section{A certain symmetry}
We see from \eqref{Mintegral} and \eqref{LLintegral} that the entangling term $\mathcal{M}$ and the \textit{mutual information} term $\mathcal{L}_{JI}$ may be written as integral transforms of the Wightman function. It is immediately clear how to write $\mathcal{L}_{JI}$ as a Fourier transform. However, $\mathcal{M}$ cannot be written as a Fourier transform of the Wightman function, because of the time-ordering restriction, $t'<t.$ Despite this,
the entangling term $\mathcal{M}$ exhibits a high degree of symmetry under interchanging the spacetime points $t,\bm{x}$ and $t',\bm{x'}$. This will allow us to find a number of other expressions for $\mathcal{M}.$

Firstly, it is clear that the integrand $\mathcal{M}(t,\bm{x},t',\bm{x'})$ is completely unchanged by exchange of primed and unprimed points, since the summands in \eqref{Mintegrand}, $\mathcal{L}_A(t,\bm{x})\mathcal{L}_B(t',\bm{x'})$ and $\mathcal{L}_A(t',\bm{x'})\mathcal{L}_B(t,\bm{x})$ transform into each other. A similar statement can be made about the volume element $\sqrt{g(t,\bm{x})g(t',\bm{x'})}$. However, there are two impediments to symmetry: namely, the integration limit forcing $t'<t$, and the Wightman function $W(t,\bm{x},t',\bm{x'}).$

Let us first examine the Wightman function. The Wightman function is conjugate-symmetric upon exchanging its two inputs; therefore, its real part is symmetric, while its imaginary part is antisymmetric. Let us then write out its division into parts:
\begin{equation}
    W(t,\bm{x},t',\bm{x'})=\Re\left[W(t,\bm{x},t',\bm{x'})\right] + \ii\Im\left[W(t,\bm{x},t',\bm{x'})\right].
\end{equation}
Now, since each part has a definite symmetry under exchange, we can re-express the integral for $t' \in (-\infty,\infty).$ Let us define a sign function $\varepsilon(\tau)=2\Theta(\tau)-1$, where $\Theta(\tau)$ is the Heaviside switching function. Then, splitting the Wightman function, we can write  
\begin{align}
\mathcal{M}=&-\lambda_A \lambda_B \frac{1}{2}\int_{-\infty}^{\infty} \dd t \int_{-\infty}^{\infty}\dd t' \int \dd^n\!\bm{x} \int \dd^n\!\bm{x'} \nonumber\\ &\times \sqrt{g(t,\bm{x})g(t',\bm{x'})} \mathcal{M}(t,\bm{x},t',\bm{x'})\nonumber\\
&\times\left(\Re\left[W(t,\bm{x},t',\bm{x'})\right]+\ii\varepsilon(t-t')\Im\left[W(t,\bm{x},t',\bm{x'})\right]\right)
\label{ReImintegral}
\end{align}
However, this particular permutation of the Wightman function is identified with the Feynman propagator $G_F$. Specifically,
\begin{align}
    \ii G_F(t,\bm{x},t',\bm{x'})&=\Re\left[W(t,\bm{x},t',\bm{x'})\right]\nonumber\\
    &+\ii\varepsilon(t-t')\Im\left[W(t,\bm{x},t',\bm{x'})\right]
\end{align}
yielding our first result:
\begin{align}
\mathcal{M}=&-\lambda_A \lambda_B \frac{1}{2}\int_{-\infty}^{\infty} \dd t \int_{-\infty}^{\infty}\dd t' \int \dd^n\!\bm{x} \int \dd^n\!\bm{x'} \nonumber\\ &\times \sqrt{g(t,\bm{x})g(t',\bm{x'})} \mathcal{M}(t,\bm{x},t',\bm{x'})\ii G_F(t,\bm{x},t',\bm{x'}).
\end{align}
This may be of some use to calculations of entanglement harvesting, since an analytic form of the Feynman propagator is often available.  This form also supports the Feynman diagram formalism introduced in \cite{Hummer:2015xaa}; it may be possible to generalize this result to higher order diagrams, and higher spin fields.

 While having this information can be useful for calculations, the Feynman propagator may not always be the most convenient tool for performing direct calculations due to the need for regularization. To find our next expression, we note that the real and imaginary part of the Wightman function are identified with the anti-commutator and commutator of the field, respectively. To be more precise,  if we define
\begin{align} 
\mathcal{C}^+(t,\bm{x},t',\bm{x'})&\coloneqq \Big\langle[\hat\Phi(t,\bm x),\hat\Phi(t',\bm x') ]_{_+}\Big\rangle_{\hat\rho(t_0)},\\
    \ii\mathcal{C}^-(t,\bm{x},t',\bm{x'})&\coloneqq \Big\langle[\hat\Phi(t,\bm x),\hat\Phi(t',\bm x') ]_{_-}\Big\rangle_{\hat\rho(t_0)},
\end{align}
where $[,]_{_-}$ and  $[,]_{_+}$ denote the commutator and anti-commutator respectively, then
\begin{align}
    \mathcal{C}^+(t,\bm{x},t',\bm{x'})&=2\Re\left[W(t,\bm{x},t',\bm{x'})\right],\\
   \mathcal{C}^-(t,\bm{x},t',\bm{x'})&=2\Im\left[W(t,\bm{x},t',\bm{x'})\right].
\end{align}
Of course, the anticommutator is symmetric, while the commutator is antisymmetric. 

Re-expressing \eqref{ReImintegral} in these terms yields our second result:
\begin{align}
\mathcal{M}=&-\lambda_A \lambda_B \frac{1}{4}\int_{-\infty}^{\infty} \dd t \int_{-\infty}^{\infty}\dd t' \int \dd^n\!\bm{x} \int \dd^n\!\bm{x'} \nonumber\\ &\times \sqrt{g(t,\bm{x})g(t',\bm{x'})} \mathcal{M}(t,\bm{x},t',\bm{x'})\nonumber\\
&\times\left(\mathcal{C}^+(t,\bm{x},t',\bm{x'})+\ii\varepsilon(t-t')\mathcal{C}^-(t,\bm{x},t',\bm{x'})\right).
\end{align}
This form of the integral is suggestive.  Since microcausality imposes that the commutator to vanish outside the light cone, one could divide the entangling term into two parts: a signalling contribution from the commutator, and a contribution from the anti-commutator. 
  Note that at leading order, all signalling comes from the commutator \cite{Martin-Martinez:2015psa}; the anti-commutator part is therefore nonsignalling.  Since the $ \mathcal{C}^+$ part does not involve signalling, one may conclude that it is harvested from the quantum state of the field.

There is another way to  recast \eqref{ReImintegral}, this time in terms of the Wightman function. Noting once again that $\mathcal{C}^-$ is antisymmetric, we can simply add it to the integral, as it will vanish after integration.
\begin{align}
\mathcal{M}=&-\lambda_A \lambda_B \frac{1}{4}\int_{-\infty}^{\infty} \dd t \int_{-\infty}^{\infty}\dd t' \int \dd^n\!\bm{x} \int \dd^n\!\bm{x'} \nonumber\\ &\times \sqrt{g(t,\bm{x})g(t',\bm{x'})} \mathcal{M}(t,\bm{x},t',\bm{x'})\nonumber\\
&\times\left(\mathcal{C}^+(t,\bm{x},t',\bm{x'})+\ii(1+\varepsilon(t-t'))\mathcal{C}^-(t,\bm{x},t',\bm{x'})\right).
\end{align}
However, since
\begin{equation}2W(t,\bm{x},t',\bm{x'})=\mathcal{C}^+(t,\bm{x},t',\bm{x'})+\ii\mathcal{C}^-(t,\bm{x},t',\bm{x'}),\end{equation} we are left with
\begin{align}
\mathcal{M}=&-\lambda_A \lambda_B \frac{1}{2}\int_{-\infty}^{\infty} \dd t \int_{-\infty}^{\infty}\dd t' \int \dd^n\!\bm{x} \int \dd^n\!\bm{x'} \nonumber\\ &\times \sqrt{g(t,\bm{x})g(t',\bm{x'})} \mathcal{M}(t,\bm{x},t',\bm{x'})\nonumber\\
&\times\left(W(t,\bm{x},t',\bm{x'})+\frac{\ii}{2}\varepsilon(t-t')\mathcal{C}^-(t,\bm{x},t',\bm{x'})\right).
\end{align}
Consider the special case where the commutator part vanishes; that is, where the two detectors have no signalling contribution. In that case, the only contribution to $\mathcal{M}$ can be re-expressed as an integral over all $t,t'$ involving the Wightman function.  This extension of the domain of integration when the commutator vanishes in the support of the switching and smearing function of the detector is at the core of the simplification obtained here.

The non-signalling part of $\mathcal{M}$ has a special connection to a previous expression. Defining  
\begin{align}
    \mathcal{M}^+=&-\lambda_A \lambda_B \frac{1}{2}\int_{-\infty}^{\infty} \dd t \int_{-\infty}^{\infty}\dd t' \int \dd^n\!\bm{x} \int \dd^n\!\bm{x'} \nonumber\\ &\times \sqrt{g(t,\bm{x})g(t',\bm{x'})} \mathcal{M}(t,\bm{x},t',\bm{x'})W(t,\bm{x},t',\bm{x'}),
    \label{Mplusintegral}\\
    \mathcal{M}^-=&-\lambda_A \lambda_B \frac{1}{4}\int_{-\infty}^{\infty} \dd t \int_{-\infty}^{\infty}\dd t' \varepsilon(t-t') \int \dd^n\!\bm{x} \int \dd^n\!\bm{x'} \nonumber\\ &\times \sqrt{g(t,\bm{x})g(t',\bm{x'})} \mathcal{M}(t,\bm{x},t',\bm{x'})\ii\mathcal{C}^-(t,\bm{x},t',\bm{x'}) 
    \label{Mminusintegral}
\end{align}
so that $\mathcal{M}=\mathcal{M}^++\mathcal{M}^-$, in the special case where the commutator vanishes between detectors, $\mathcal{M}^-$ vanishes, and thus $\mathcal{M}=\mathcal{M}^+$.
Note that, by the previous arguments, this expression is unchanged if we replace $W$ by $\mathcal{C}^+/2$, hence the name.

Now, let us examine more closely the $\mathcal{L}_I(t,\bm{x})$ terms in \eqref{Lintegrand}. Since all the other parts are real we have 
\begin{equation}
    \mathcal{L}^*_I(\Omega;t,\bm{x})=\mathcal{L}_I(-\Omega;t,\bm{x}) 
\end{equation}
or in other words, 
the effect of taking the complex conjugate is to replace $\Omega_I$ by $-\Omega_I.$  
We can then rewrite \eqref{Mintegrand} in the suggestive form 
\begin{align}
    \mathcal{M}(t,\bm{x},t',\bm{x'})&=\mathcal{L}_A(\Omega_A;t,\bm{x})\mathcal{L}^*_B(-\Omega_B;t',\bm{x'})\nonumber\\
    &+\mathcal{L}_B(\Omega_B;t,\bm{x})\mathcal{L}^*_A(-\Omega_A;t',\bm{x'})
\end{align}
thereby expressing \eqref{Mplusintegral} in terms of \eqref{LLintegral} with modified gaps. Indicating dependence on detector gaps by $\mathcal{L}_{IJ}[\Omega_I,\Omega_J],$  we have therefore found 
\begin{equation}\label{central}
    \mathcal{M}^+=-\frac{1}{2}\left(\mathcal{L}_{AB}[\Omega_A,-\Omega_B]+\mathcal{L}_{BA}[\Omega_B,-\Omega_A]\right)
\end{equation}
which is the central result of this paper.

Put another way, since $\mathcal{M}=\mathcal{M}^++\mathcal{M}^-,$
\begin{align}
    \mathcal{M}=&-\frac{1}{2}\left(\mathcal{L}_{AB}[\Omega_A,-\Omega_B]+\mathcal{L}_{BA}[\Omega_B,-\Omega_A]\right)\nonumber\\
    &-\lambda_A \lambda_B \frac{1}{4}\int_{-\infty}^{\infty} \dd t \int_{-\infty}^{\infty}\dd t' \varepsilon(t-t') \int \dd^n\!\bm{x} \int \dd^n\!\bm{x'} \nonumber\\ &\times \sqrt{g(t,\bm{x})g(t',\bm{x'})} \mathcal{M}(t,\bm{x},t',\bm{x'})\ii\mathcal{C}^-(t,\bm{x},t',\bm{x'})\; .
\label{finalres}    
\end{align}
When $\mathcal{C}^-(t,\bm{x},t',\bm{x'})$ vanishes over the support of the switching  and smearing functions, $\mathcal{M}^-$ vanishes, and thus the entangling term can then be expressed in terms of the mutual information term, with different gaps.

The expression \eqref{central} (or alternatively \eqref{finalres}) is our main result. It indicates that in the special case where the commutator vanishes, the $\mathcal{M}$ term can be expressed in terms of the $\mathcal{L}_{JI}$ term, albeit with asymmetric values of $\Omega_I$.  As we will see in the example below, this allow a re-expression of  the entanglement term of the density matrix of the two detectors in terms of the calculationally much simpler \textit{mutual information} terms.  Importantly, this is always true if the detectors are spacelike separated, a case of great theoretical interest  in entanglement harvesting.; in the particular spacetimes where   Huygens principle holds,  such as (3+1)-D Minkowski spacetime, or conformally coupled fields in FRW \cite{Ellis,czapor,McLenaghan,Sonego:1991sq,Comm2,Blasco:2015eya}, it also is true if the detectors are \textit{timelike} separated.

 Furthermore, the result \eqref{finalres} also indicates precisely how $\mathcal{M}$ depends on the commutator; in some sense, it is possible to precisely quantify how much of the entangling term is due to signalling between A and B.  Since $\mathcal{L}_{JI}$ can be quickly expressed in terms of Fourier transforms, we expect this result will be a boon to researchers studying entanglement harvesting, especially in the spacelike case.

There is another intriguing implication to this expression. Previous examinations of the causality of the Unruh-DeWitt model have noted that UV-cutoffs can introduce causality violations into the model. (For further information, refer to e.g. \cite{Martin-Martinez:2015psa}.) However, expressing the entangling term in this way helps us understand why this happens. Since the $\mathcal M^-$ term contains the contributions of the commutator, invoking a UV limit has the effect of `blurring' the commutator, leading to contributions outside the lightcone. Calculating the $\mathcal M^-$ term separately may thus limit the damage done by taking this UV limit. For instance, in Minkowski space in four dimensions, the strong Huygens principle guarantees that the commutator term is supported on the light cone; it is thus particularly simple to calculate $\mathcal M^-$ exactly. 
On the other hand, it also implies something interesting about $\mathcal M^+$: namely, since this integral is completely unable to distinguish a reversal of time ordering, it \textit{cannot} contribute to signaling. It would be interesting to assess in whether this quality holds when UV cutoffs are applied.

As a brief demonstration, suppose additionally that two pointlike detectors remain stationary in a space with a Killing time. If the field state is a vacuum, we may take the mode expansion of the Wightman function:
\begin{equation}
    W(t,\bm{x},t',\bm{x'})=\sum_{nlm}\frac{1}{2\omega_{nlm}}e^{-\ii\omega_{nlm} (t-t')}\varphi_{nlm}(x)\overline{\varphi}_{nlm}(x').
\end{equation}
Note that this expression implicitly contains the commutator in its imaginary part; however, faithfully representing its compact support requires continuing the sum through the UV limit. In this case, we may express \eqref{Mplusintegral} with respect to Fourier transforms of the switching functions:
\begin{widetext}
\begin{align}
    \mathcal{M}^+=-\frac{1}{2}\lambda_B\lambda_A \sum_{n l m}\frac{\pi}{\omega}&\left(\hat{\chi}_B \left(\Omega_B-\frac{dt}{\dd\tau_B}\omega\right)\hat{\chi}_A\left(\Omega_A+\frac{dt}{\dd\tau_A}\omega\right)\varphi_{n  l m}(x_B)\varphi^*_{\omega l m}(x_A)\right.\nonumber\\
    &+\left.\hat{\chi}_A\left(\Omega_A-\frac{dt}{\dd\tau_A}\omega\right)\hat{\chi}_B\left(\Omega_B+\frac{dt}{\dd\tau_B}\omega\right)\varphi_{n  l m}(x_A)\varphi^*_{\omega l m}(x_B)\right).\label{symmetricM}
\end{align}
\end{widetext}
This compares quite well to the resulting integral in \eqref{LLintegral}:
\begin{align}
\mathcal{L}_{IJ}=&\lambda_J\lambda_I \sum_{\omega lm}\frac{\pi}{\omega}
\varphi_{n  lm}(x_J)\varphi^*_{\omega lm}(x_I)\nonumber\\
&\times\hat{\chi}_{I}\left(\omega\frac{dt}{\dd\tau_I}+\Omega_I\right)\hat{\chi}^*_{J}\left(\omega\frac{dt}{\dd\tau_J}+\Omega_J\right).
\label{Lnm}
\end{align}
While the integral \eqref{Mminusintegral} remains, the fact that the commutator is supported on and within the light cone typically implies a great deal of computational savings. In fact, in spacetimes where the strong Huygens principle holds, not only does this have the effect of reducing the two time integrals to one, but this typically allows us to evaluate the remaining time integral analytically. We may thus be assured of the causality of our calculation.

We emphasize that our results apply very generally: the detectors may possess different trajectories, have different gaps, and be switched differently, as long as the Wightman function itself has the symmetry we need: namely, that it is conjugate symmetric. In the special case where the imaginary part of the Wightman function is associated with the commutator, as in the free field vacuum state, it is even possible to express the entangling part in terms of other terms in the density matrix, when the commutator vanishes. Since these terms are typically much easier to calculate, we have thus found a way to greatly reduce the cost of computing the entangling term, in the most general case.

\section{An example in Minkowski space}

In order to verify our result, let us examine previous work done in Minkowski space \cite{Pozas-Kerstjens:2015gta}. Let us take two identical detectors with switching functions:
\begin{equation}
    \chi_I(t)=e^{-(t-t_I)^2/T^2},
\end{equation}
where $T$ is the switching time.


There is a small difficulty: the expression for $\mathcal L_{AB}$ in \cite{Pozas-Kerstjens:2015gta} assumes that both detectors have the same gap. If this is not the case, then not only does $\mathcal L_{AB}$ depend on $t_B-t_A,$ but it will also depend on $t_B+t_A$. While this is typical for calculations where detectors have unequal gaps, it does mean we must proceed with caution. Note that they define $L_I$ differently: for this section only we will adopt their conventions.

According to their notation, let us take $t_A=-t_B=-t_0/2$; that is, detector A switches first, and we divide the delays equally in coordinate time (or, more specifically, in ``mode time"). Let us use their notation for $G_1(\kappa,\tau_\mu, \alpha),$ adding in the unitless gap as an additional parameter. In that case, we find:
\begin{align}
    \mathcal{M}^+&=-\frac{1}{2}\left(\mathcal{L}_{AB}[\alpha,-\alpha]+\mathcal{L}_{BA}[\alpha,-\alpha]\right)\\
    &=-\frac{\lambda^2}{8\pi^2T^2\beta}\int_0^\infty d|\bm{\kappa}|\sin(\bm{\kappa}\beta)e^{-\frac{1}{2}\bm{\kappa}^2\delta^2}\nonumber\\
    &\times\left(G_1(\bm{\kappa},-\gamma/2,\alpha)G_1(\bm{\kappa},\gamma/2,-\alpha)^*\right.\nonumber\\
    &+\left.G_1(\bm{\kappa},\gamma/2,\alpha)G_1(\bm{\kappa},-\gamma/2,-\alpha)^*\right).
\end{align}
We then evaluate the $G_1$ products. Again, note that our choice of delays suppresses dependence on $t_A+t_B$.
\begin{align}
    G_1&(\bm{\kappa},-\gamma/2,\alpha)G_1(\bm{\kappa},\gamma/2,-\alpha)^*\nonumber\\
    &=\pi T^2 e^{\frac{1}{2}(|\bm{\kappa}|^2+\alpha^2)}e^{-\ii|\bm{\kappa}|\gamma}\\
    G_1&(\bm{\kappa},\gamma/2,\alpha)G_1(\bm{\kappa},-\gamma/2,-\alpha)^*\nonumber\\
    &=\pi T^2 e^{\frac{1}{2}(|\bm{\kappa}|^2+\alpha^2)}e^{+\ii|\bm{\kappa}|\gamma}
\end{align}
Thus, our integral becomes
\begin{align}
    \mathcal{M}^+&=-\frac{\lambda^2 e^{-\frac{1}{2}\alpha^2}}{4\pi\beta}\int_0^\infty d|\bm{\kappa}|\sin(\bm{\kappa}\beta)\cos(\bm{\kappa}\gamma)e^{-\frac{1}{2}\bm{\kappa}^2(\delta^2+1)}\\
    &=-\frac{\lambda^2 e^{\frac{1}{2}\alpha^2}}{8\sqrt{2\pi}\beta\sqrt{1+\delta^2}}\nonumber\\
    &\times \left(e^{-\frac{(\beta-\gamma)^2}{2(1+\delta^2)}}\textrm{erfi}\left(\frac{\beta-\gamma}{\sqrt{2}\sqrt{1+\delta^2}}\right)\right.\nonumber\\
    &+\left.e^{-\frac{(\beta+\gamma)^2}{2(1+\delta^2)}}\textrm{erfi}\left(\frac{\beta+\gamma}{\sqrt{2}\sqrt{1+\delta^2}}\right)\right)
\end{align}

Amusingly, it appears our result for $\mathcal{M}^+$ \textit{almost} appears in \cite{Pozas-Kerstjens:2015gta}. Specifically, this quantity is very close to their value for $|\mathcal{M}_{non}|,$ since $\textrm{erfi}(z)=-\ii \textrm{erf}(\ii z).$ In fact, rewriting their value highlights what the difference is. Multiplying the quantity inside the absolute value by $-\ii$, we find:
\begin{align}
    |\mathcal{M}_{non}|&=\frac{\lambda^2 e^{-\frac{1}{2}\alpha^2}}{8\sqrt{2\pi}\beta\sqrt{1+\delta^2}}\nonumber\\
    &\times \left| e^{-\frac{(\beta-\gamma)^2}{2(1+\delta^2)}}\left[\textrm{erfi}\left(\frac{\beta-\gamma}{\sqrt{2}\sqrt{1+\delta^2}}\right)-\ii\right]\right.\nonumber\\
    &+\left.e^{-\frac{(\beta+\gamma)^2}{2(1+\delta^2)}}\left[\textrm{erfi}\left(\frac{\beta+\gamma}{\sqrt{2}\sqrt{1+\delta^2}}\right)+\ii\right]\right|. \label{Mnonintegral}
\end{align}
The only difference appears to be an imaginary term, which is compared to the imaginary error function. Recall that  $|\mathcal{M}|\approx|\mathcal{M}_{non}|$ in the case where $\gamma \gg 1.$ If we additionally assume $\gamma/\sqrt{1+\delta^2} \gg \beta/\sqrt{1+\delta^2}$ \textit{or} $\gamma/\sqrt{1+\delta^2} \ll \beta/\sqrt{1+\delta^2}$, this difference becomes insignificant. This corresponds precisely to the cases where the detectors are timelike or spacelike separated respectively, with the commutators vanishing.

As for the commutator part, things get rather more complicated. In Minkowski (3+1)-space, the commutator $\ii\mathcal{C}^-=[\hat\Phi(x^\mu),\hat\Phi(x'^\nu)]$ may be written as
\begin{equation}
    \ii\mathcal{C}^-=\frac{\ii}{4\pi|\bm{x}-\bm{x'}|}(\delta(t-t'+|\bm{x}-\bm{x'}|)-\delta(t-t'-|\bm{x}-\bm{x'}|)).
\end{equation}
Of course, since the commutator is supported on the light cone, it is a uniquely UV feature; any UV cutoff will `blur' the commutator, and cause causal issues \cite{Martin-Martinez:2015psa}. Once again, we set $t_A+t_B=0.$ In order to increase the symmetry of the result, after integrating the delta function, we translate $t$.
Under these conditions, we may write
\begin{align}
    \mathcal{M}^-&=\frac{\ii\lambda^2}{8\pi} \int_{-\infty}^\infty \dd\tau \int \dd^3\bm{\beta} \int \dd^3\bm{\beta'}\frac{e^{-2\tau^2+2\ii\alpha\tau}}{|\bm{\beta}-\bm{\beta'}|}\nonumber\\
    &\left(e^{-\frac{1}{2}(|\bm{\beta}-\bm{\beta'}|+\gamma)^2} F_A(\bm{x})F_B(\bm{x'}) \right.\nonumber\\
    &\left.+ e^{-\frac{1}{2}(|\bm{\beta}-\bm{\beta'}|-\gamma)^2} F_B(\bm{x})F_A(\bm{x'})\right)
\end{align}

If we additionally assume the detectors are pointlike, setting $\delta=0,$ we can do this integral analytically, resulting in
\begin{align}
    \mathcal{M}^-&=\frac{\ii\lambda^2e^{-\frac{1}{2}\alpha^2}}{8\sqrt{2\pi}\beta}\left(e^{-\frac{(\beta+\gamma)^2}{2}}+ e^{-\frac{(\beta-\gamma)^2}{2}} \right).
\end{align}
Therefore, our expression for $\mathcal{M}$ with $\delta=0$ is simply
\begin{align}
    \mathcal{M}&=-\frac{\lambda^2e^{-\frac{1}{2}\alpha^2}}{8\sqrt{2\pi}\beta}\nonumber\\
    &\times \left( e^{-\frac{(\beta-\gamma)^2}{2}}\left[\textrm{erfi}\left(\frac{\beta-\gamma}{\sqrt{2}}\right)-\ii\right]\right.\nonumber\\
    &+\left.e^{-\frac{(\beta+\gamma)^2}{2}}\left[\textrm{erfi}\left(\frac{\beta+\gamma}{\sqrt{2}}\right)-\ii\right]\right) \label{Mpmresult}
\end{align}
This new expression is even closer to $\mathcal{M}_{non}$. Not precisely, because even if we set $\delta=0,$ the imaginary summands end up with the same sign, rather than the opposite sign as in \eqref{Mnonintegral}. The difference is negligible unless both $\beta$ and $\gamma$ are small, which is consistent with the conditions with which \eqref{Mnonintegral} was derived.

\begin{figure}[hbtp]
    \centering
    \includegraphics[width=0.95\columnwidth,trim={0 0 0 0},clip]{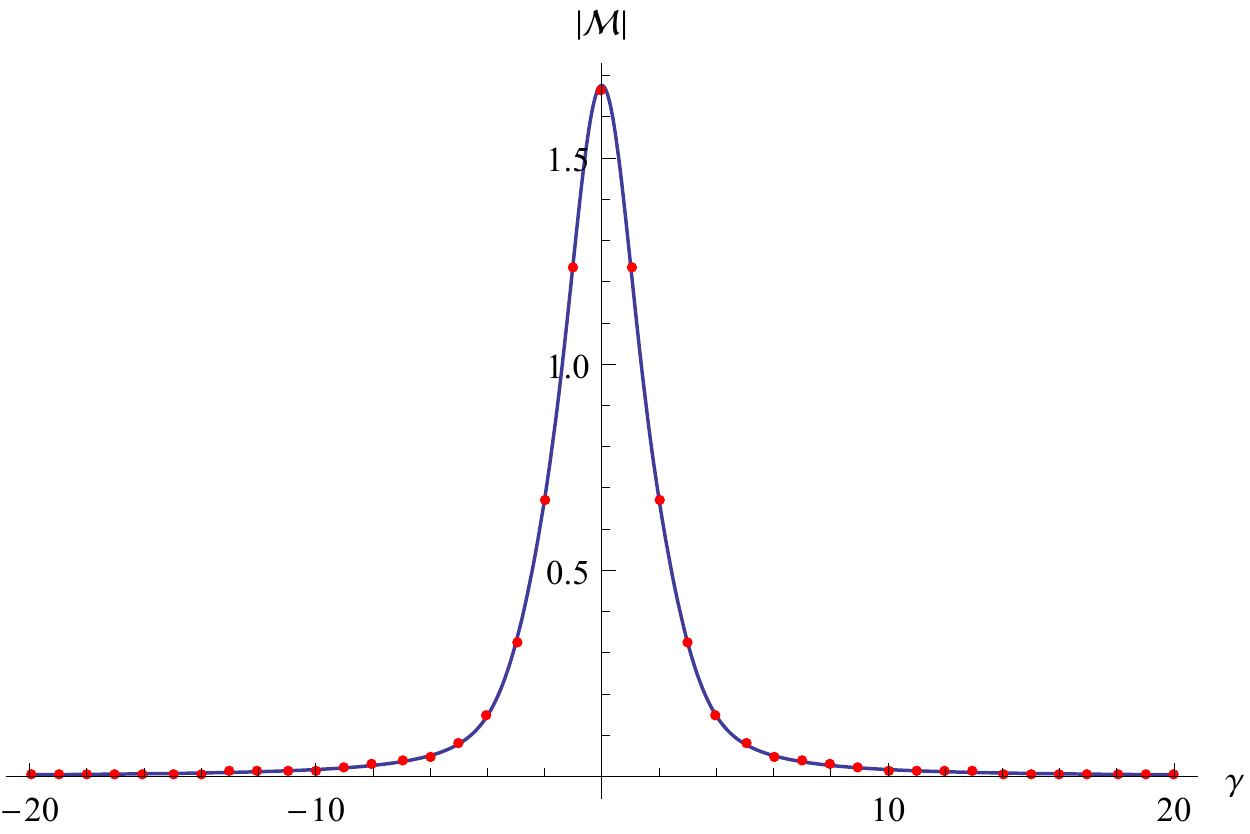}
    \caption{A plot of our expression for $\mathcal{M}$ (blue line) against the results in \cite{Pozas2016} (red points) for $\beta=1$.}
    \label{fig:pozas1}
\end{figure}

\begin{figure}[hbtp]
    \centering
    \includegraphics[width=0.95\columnwidth,trim={0 0 0 0},clip]{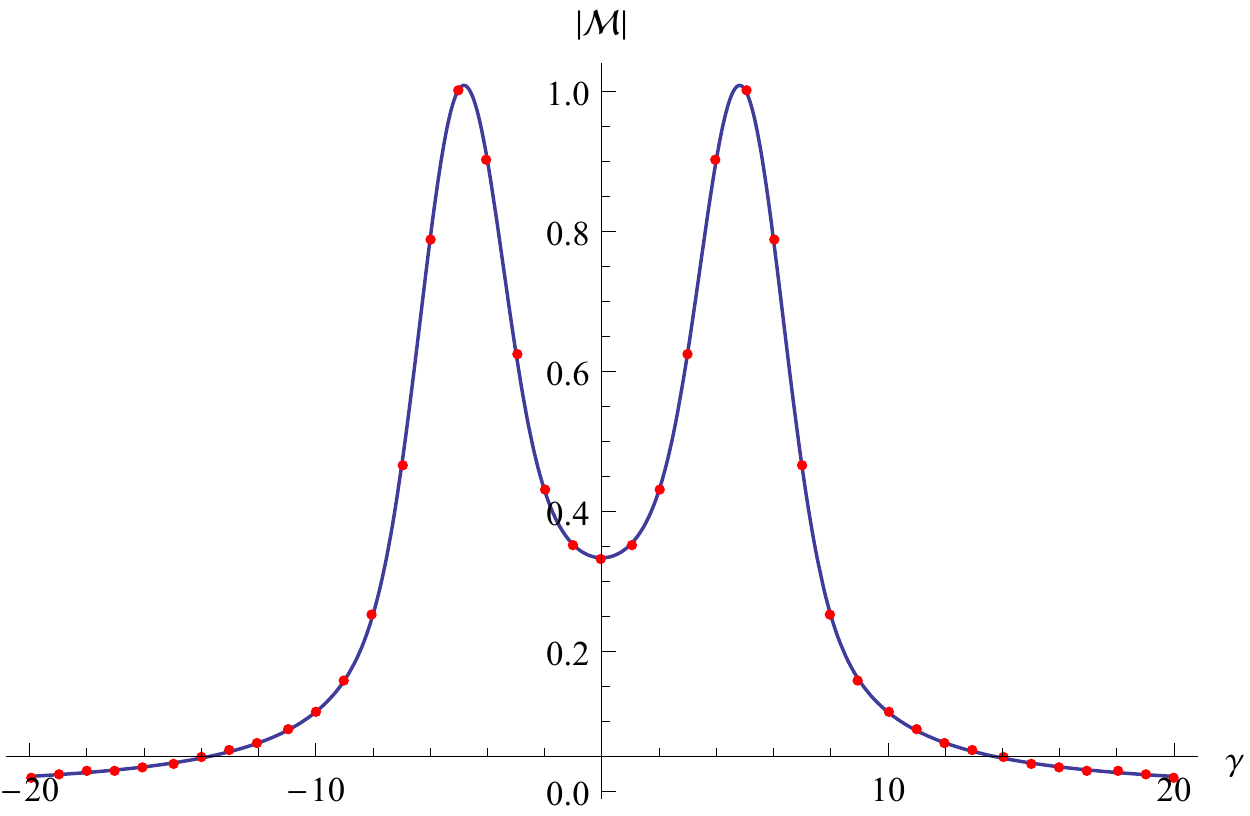}
    \caption{A plot of our expression for $\mathcal{M}$ (blue line) against the results in \cite{Pozas2016} (red points) for $\beta=5$.}
    \label{fig:pozas5}
\end{figure}

\begin{figure}[hbtp]
    \centering
    \includegraphics[width=0.95\columnwidth,trim={0 0 0 0},clip]{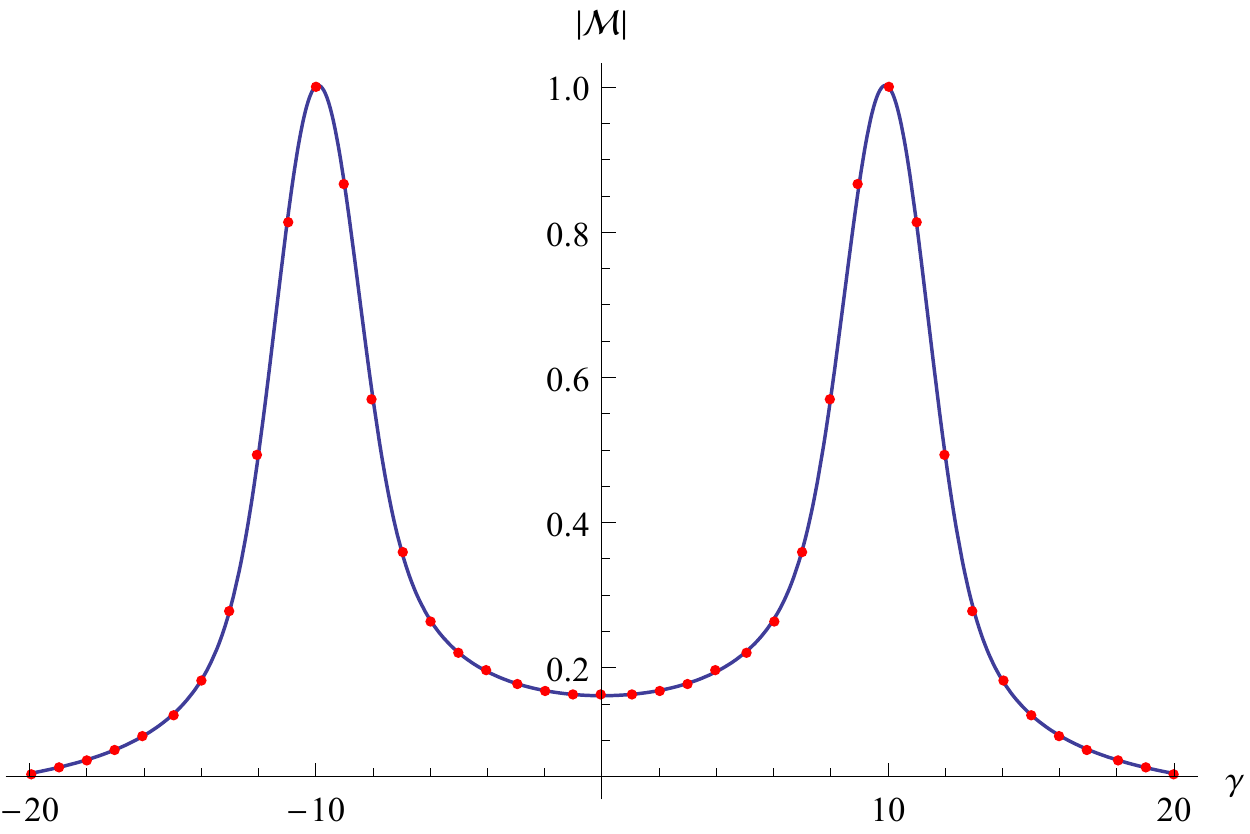}
    \caption{A plot of our expression for $\mathcal{M}$ (blue line) against the results in \cite{Pozas2016} (red points) for $\beta=10$.}
    \label{fig:pozas10}
\end{figure}

Even where an analytic integral does not exist for $\mathcal{M},$ we can still pick parameters and plot things numerically. Amusingly, both our expression \ref{Mpmresult} for $\mathcal{M}$ and that of \cite{Pozas-Kerstjens:2015gta}, equation (35), have $\alpha$ factor out of the final integral, and depend on it in the same way; therefore, $\alpha$ merely acts as a scale factor for the purposes of this comparison. Therefore, after dividing out a common factor $\frac{\lambda^2 e^{-\alpha^2/2}}{8\sqrt{2\pi}\beta}$, we plotted both expressions against $\gamma$, for $\delta=0$ and $\beta = 1,5,10$ in Fig. \ref{fig:pozas1}, \ref{fig:pozas5}, \ref{fig:pozas10} respectively. We believe that the agreement of numerical results to our expression is satisfactory; however, we will leave the question of whether our expression \textit{is} the analytic integral of the expression in \cite{Pozas-Kerstjens:2015gta} to another day:
\begin{equation}
    E(\bm{\kappa},\gamma)=e^{\ii\gamma|\bm{\kappa}|}\textrm{erfc}\left(\frac{\gamma+\ii|\bm{\kappa}|}{\sqrt{2}}\right),
\end{equation}
\begin{align}
    \sqrt{\frac{2}{\pi}}&\int_{0}^\infty d|\bm{\kappa}|\,\sin(\beta|\bm{\kappa}|)e^{-\bm{\kappa}^2/2}(E(\bm{\kappa},\gamma)+E(\bm{\kappa},-\gamma))\nonumber\\
    &\approx e^{-\frac{(\beta-\gamma)^2}{2}}\left[\textrm{erfi}\left(\frac{\beta-\gamma}{\sqrt{2}}\right)-\ii\right]\nonumber\\
    &+e^{-\frac{(\beta+\gamma)^2}{2}}\left[\textrm{erfi}\left(\frac{\beta+\gamma}{\sqrt{2}}\right)-\ii\right].
\end{align}

\begin{figure}[hbtp]
    \centering
    \includegraphics[width=0.95\columnwidth,trim={0 0 0 0},clip]{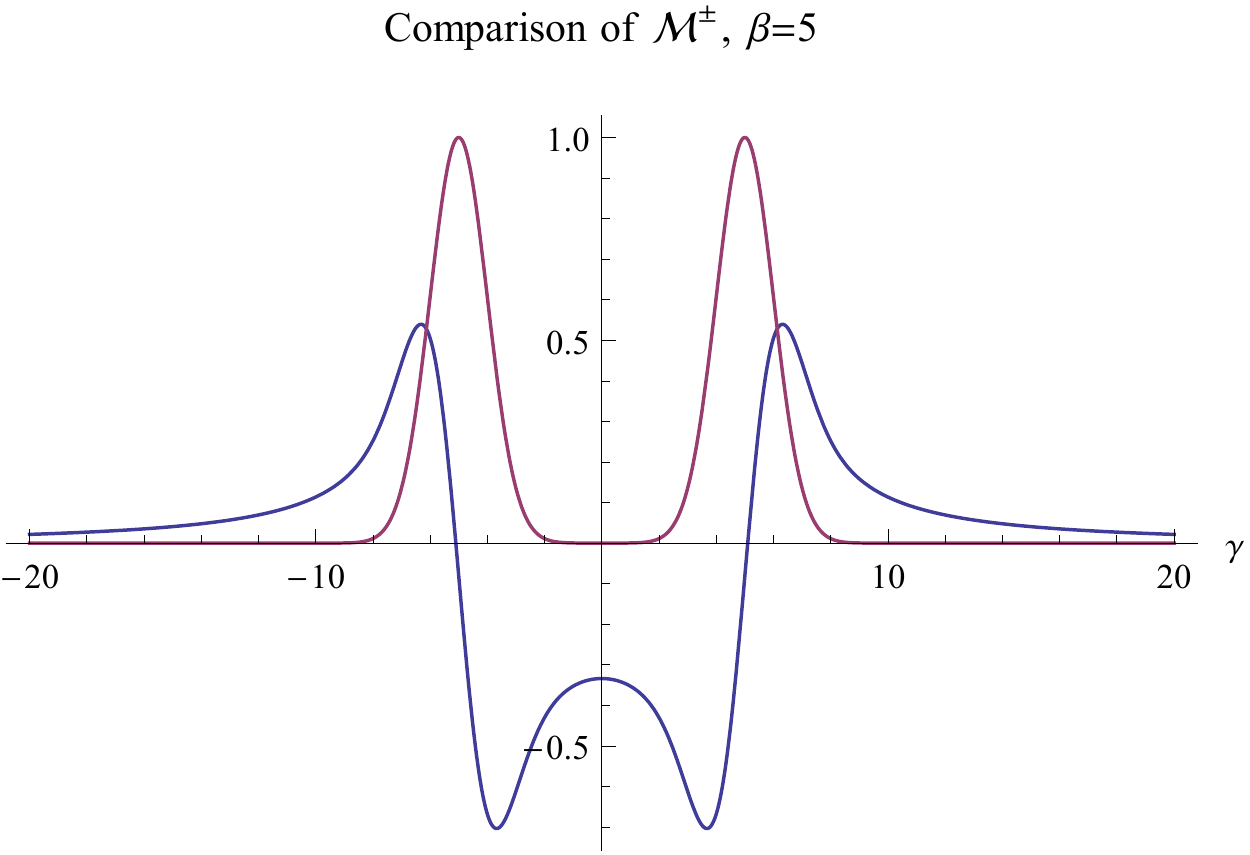}
    \caption{Comparing $\mathcal{M}^\pm$ for $\beta=5$: $\mathcal M^+$ solid blue, $\mathcal M^-/\ii$ dashed orange. Note that $\mathcal{M}^+$ appears to be smallest in magnitude on the light cone, where $\mathcal{M}^-$ is largest.}
     \label{fig:Mpm5}
\end{figure}

Finally, in Fig. \ref{fig:Mpm5} we also plot $\mathcal{M}^\pm$ against each other for $\beta=5,$ in order to better visualize their dependence on space and time.
This plot makes a rather counterintuitive feature of $\mathcal{M}^+$ clear: namely, it seems to cross zero as $\mathcal{M}^-$ grows largest, although these two things do not occur at precisely the same values. Thus, in some sense, $\mathcal{M}^+$ seems to abhor the light cone. It is for this reason that we might call $\mathcal{M}^+$ the `non-communicating' part of $\mathcal{M}$. Whether this vanishing of $\mathcal{M}^+$ is a general phenomenon is a matter for further research.  This graph also demonstrates exactly how much the communicating term $\mathcal{M}^-$ contributes to the entanglement; it is gratifying to see that in the spacelike region $\mathcal{M}^+$ does indeed dominate. Note that even when $\tau=0$, $\mathcal{M}^-$ does \textit{not} vanish: $\Im\mathcal{M}$ is indeed positive for all values of $\tau$.

\section{Conclusions}

In many quantum field systems, in both flat and curved spacetimes, the Wightman function is only known with respect to a mode expansion. Calculating  multiple detector statistics in such contexts thus requires taking {two} or more nested integrals, which leads to a high degree of computational complexity  and makes it difficult to find closed expressions. Additionally, preservation of causality requires careful consideration of the UV limit; it has been shown \cite{Martin-Martinez:2015psa} that taking the UV limit of the integrals involved can lead to violations of causality. 

We found that both of these features can be explained in terms of the commutator of the field, which corresponds to the imaginary part of the Wightman function. In the cases where the commutator vanishes, the  nested time integrals may be transformed into Fourier transforms, and thus great computational savings may be achieved, with little risk of causality violation. We also found that the full expression, Fourier plus commutator, allows for a very accurate calculation of the entangling term with much lower computational cost. We have demonstrated this for the Minkowski space case previously analyzed in \cite{Pozas-Kerstjens:2015gta}, and found a solution in  closed-form where none was previously known.

Our analysis answers some questions about the causality of the detector model, and raises others. The ability to isolate the commutator part of the entangling term, $\mathcal{M}^-$, explains why applying a UV limit can cause causality violations. However, the converse remains to be shown: namely, that the anti-commutator part $\mathcal{M}^+$ does not. We would also like to extend this result to even more generality: higher order interactions, higher spin fields, and higher order couplings to the field. Even in its current form, however, we expect that our results may be of great practical use to the entanglement harvesting community, especially in cases where the Wightman function is expressed as a mode sum, such as the vacuum states of fields on curved space: we expect to publish more results in this vein in the near future. 

\section*{Acknowledgements} 
E. M-M. acknowledges the support of the Ontario Early Researcher Award. R.B.M and E. M-M. acknowledges the support of the NSERC Discovery program.

\bibliography{MLpaper}

\end{document}